\documentclass[pre, amsmath]{revtex4-2}

\usepackage[dvipdfmx]{graphicx}
\usepackage{siunitx}

\begin{document}

\title{Theoretical analysis of the maximum range of a projectile released from a pendulum}
\author{Ken Yamamoto}
\affiliation{Faculty of Science, University of the Ryukyus, Nishihara, Okinawa 903--0213, Japan}

\begin{abstract}
The motion of a projectile released from a simple pendulum is analyzed, with particular emphasis on investigating the optimal release angle that maximizes the horizontal range and the corresponding maximum range.
This system serves as a simplified model of the Tarzan jump problem.
Using simple analytical methods, the optimal release angle is shown to be characterized by a cubic equation and to increase with the initial velocity.
In addition, asymptotic expressions for both the optimal angle and maximum range are derived in the limits of low and high initial velocity.
\end{abstract}

\maketitle

\section{Introduction}
Throwing or launching an object over a long distance and landing it at a desired point are fundamental problems in mechanics, arising in contexts such as throwing events, ball sports, and the firing of artillery shells and rockets.
The analysis of projectile motion appears even at an early stage in physics.
When air resistance is negligible, it is well known that a projectile follows a parabolic trajectory and that a launch angle of $\pi/4\,\mathrm{rad} (=\ang{45})$ from the horizontal yields the maximum range when the projectile is released from ground level.

A variety of projectile motion problems have been analyzed to date.
For example, when an object is released from a height $H$ above the ground, the angle
\begin{equation}
\theta=\arctan\frac{v_0}{\sqrt{v_0^2+2gH}}
\label{eq:thetamax_simple}
\end{equation}
yields the maximum range
\begin{equation}
R=\frac{v_0}{g}\sqrt{v_0^2+2gH},
\label{eq:Rmax_simple}
\end{equation}
where $v_0$ is the launch velocity and $g$ is the gravitational acceleration~\cite{Ganci2014}.
Further generalizations include air resistance proportional to velocity, for which the optimal angle and the maximum range can be expressed exactly using the Lambert W function~\cite{Hu2012}.
When air resistance is proportional to the square of the speed, the trajectory cannot be obtained exactly; however, numerical analyses~\cite{Hayen2003} and approximate solutions in limiting cases~\cite{Cohen2014} have been developed.
The motion of a projectile in complex fluids is described using fractional calculus to capture non-locality and hysteresis effects~\cite{Lazopoulos2021}.

Thus, variations of simple projectile motion, where \ang{45} is optimal, yield interesting and nontrivial properties, both physically and mathematically.
The present work focuses on the motion of an object released from a simple pendulum, commonly referred to as the ``Tarzan jump.''
This problem combines pendulum and parabolic motions, both treated in elementary mechanics.
For this aspect, this problem has been extensively examined in the physics education literature, particularly aimed at developing laboratory experiments, education projects, and instructional material~\cite{Trout2001, Bittel2005, Rave2013}.
Although some existing studies performed numerical calculations to achieve deeper understanding~\cite{Mungan2011, Shima2012}, the theoretical properties have not been fully investigated.
As stated in the following sections, this problem is more complex than it appears and exhibits an intriguing mathematical structure.

In this work, we establish a theoretical basis for this problem, particularly the characterization of the optimal release angle and maximum range.
The basic equation is a cubic equation for the optimal release angle~\cite{Bittel2005}.
We explain its derivation in detail, with careful attention to theoretical insight and computational efficiency.
We also establish that the equation is physically valid by proving the unique existence of the appropriate solution.
Systematic analysis of this equation allows to precise quantitative results, including asymptotic forms in the limits of low and high initial velocity.
Thus, the present study provides a firm theoretical foundation for the Tarzan problem and presents a detailed quantitative analysis.

\section{Formulation}\label{sec2}
\begin{figure}[tb!]\centering
\includegraphics[clip]{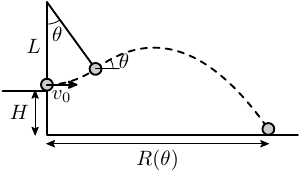}
\caption{
Problem setup.
The object starts moving horizontally with initial velocity $v_0$ at height $H$, and it swings as the bob of a pendulum of length $L$.
When the pendulum reaches an angle $\theta$, the object is released and undergoes free fall.
The horizontal distance $R(\theta)$ is measured from the initial position.
}
\label{fig1}
\end{figure}

Figure~\ref{fig1} illustrates the problem setup.
Initially, an object is located at a height $H$ above the ground, corresponding to the lowest point of a simple pendulum of length $L$.
The object begins to swing with an initial horizontal velocity $v_0$.
When the pendulum reaches an angle $\theta$ from the downward vertical, it is released and subsequently undergoes free fall, eventually landing at a horizontal distance $R(\theta)$.
This motion models Tarzan's action: he runs down a hill of height $H$, grasps a vertically hanging rope of length $L$, swings along its arc, releases the rope at an angle $\theta$, and lands on the ground.
As shown in Fig.~\ref{fig1}, $R(\theta)$ is measured horizontally from the initial position of the object (i.e., the lowest point of the pendulum), rather than from the release point.
At the moment of release, the object is assumed to move tangentially to the pendulum trajectory, with mechanical energy conserved.
Air resistance is neglected throughout the analysis.

Two approaches to specifying the initial condition of the pendulum motion have been adopted in previous studies.
One approach specifies the velocity at the lowest point of the pendulum, whereas the other specifies an initial angle $\theta_0$ from which the motion begins.
These two descriptions are related by $v_0^2=2gL(1-\cos\theta_0)$.
In the present study, we adopt the specification of $v_0$, as it allows for a broader class of motions; in particular, if $v_0>2\sqrt{gL}$, no corresponding angle $\theta_0$ exists.

This problem involves two competing effects.
A smaller release angle $\theta$ results in a higher launch speed, whereas a larger $\theta$ leads to a greater launch angle and a more forward release position.
Consequently, $R(\theta)$ is expected to attain its maximum at an intermediate value of $\theta$, neither too small nor too large.
Rave and Sayers~\cite{Rave2013} refer to this trade-off as ``Tarzan's dilemma.''

\section{Analysis}
\subsection{Characterization of maximum range and optimal angle}\label{sec3.1}
As the derivation of $R(\theta)$ has been presented in previous studies~\cite{Mungan2011, Shima2012, Bittel2005, Rave2013}, we provide only a brief outline here.
The speed of the object when the pendulum reaches an angle $\theta$ is given by $v = \sqrt{v_0^2-2gL(1-\cos\theta)}$, according to the conservation of mechanical energy.
The horizontal and vertical components of the launch velocity are $v\cos\theta$ and $v\sin\theta$, respectively.
Considering the vertical motion under a constant downward acceleration $g$, the time from release to landing is
\begin{equation}
T_\mathrm{flight}=\frac{1}{g}\left(v\sin\theta+\sqrt{v^2\sin^2\theta+2g(H+L(1-\cos\theta))}\right).
\label{eq:Tflight}
\end{equation}
By substituting the expression for $v$, the range $R(\theta)$ is obtained as
\begin{align}
R(\theta)&=L\sin\theta + T_\mathrm{flight} v\cos\theta\nonumber\\
&=L\sin\theta+\left(\frac{v_0^2}{g}-2L(1-\cos\theta)\right)\sin\theta\cos\theta\nonumber\\
&\quad+\frac{\cos\theta}{g}\sqrt{(v_0^2-2gL(1-\cos\theta))[v_0^2\sin^2\theta+2gH+2gL(1-\cos\theta)\cos^2\theta]}.
\label{eq1}
\end{align}

To simplify this expression and highlight the essential features of the problem, we introduce nondimensional quantities by scaling lengths with $L$:
\begin{equation}
\tilde{R}(\theta)=\frac{R(\theta)}{L},\quad
\tilde{H}=\frac{H}{L},
\label{eq:nondim_length}
\end{equation}
and define a nondimensional velocity using $L$ and $g$:
\begin{equation}
\tilde{v}_0=\frac{v_0}{\sqrt{2gL}}.
\label{eq:nondim_velocity}
\end{equation}
The factor $\sqrt{2}$ is introduced to maintain consistency with the notation of Mungan and Lipscombe~\cite{Mungan2011}.
A tilde denotes the nondimensional quantities in this study.
The nondimensional form of Eq.~\eqref{eq1} becomes
\begin{align}
\tilde{R}(\theta)&=\sin\theta+2(\tilde{v}_0^2-(1-\cos\theta))\sin\theta\cos\theta\nonumber\\
&\quad+2\cos\theta\sqrt{(\tilde{v}_0^2-(1-\cos\theta))[\tilde{v}_0^2\sin^2\theta+\tilde{H}+(1-\cos\theta)\cos^2\theta]}.
\label{eq:R}
\end{align}

The optimal angle $\theta_\mathrm{max}$ that maximizes $\tilde{R}(\theta)$ is determined by solving $d\tilde{R}(\theta)/d\theta=0$.
The function $\tilde{R}(\theta)$ has a complicated form containing a trigonometric polynomial within a square root.
Naive or brute-force calculations of the derivative of such a function likely give poor insight or eventually become unmanageable.
We propose the following ideas to improve transparency and avoid cumbersome arithmetic.

We introduce the following two functions:
\begin{equation}
F(\theta)=\tilde{v}_0^2-(1-\cos\theta),\quad
G(\theta)=\tilde{v}_0^2\sin^2\theta+\tilde{H}+(1-\cos\theta)\cos^2\theta=\tilde{H}+\tilde{v}_0^2-F(\theta)\cos^2\theta,
\label{eq:FGdef}
\end{equation}
whose derivatives are
\[
F'(\theta)=-\sin\theta,\quad
G'(\theta)=(2F(\theta)+\cos\theta)\sin\theta\cos\theta.
\]
The function $\tilde{R}(\theta)$ can be expressed as
\begin{equation}
\tilde{R}(\theta)=\sin\theta+2F(\theta)\sin\theta\cos\theta+2\cos\theta\sqrt{F(\theta)G(\theta)},
\label{eq:R_FG}
\end{equation}
and its derivative becomes
\begin{align*}
\frac{d\tilde{R}(\theta)}{d\theta}
&=\cos\theta+2F'(\theta)\cos\theta\sin\theta+2F(\theta)(\cos^2\theta-\sin^2\theta)\\
&\quad+\frac{F'(\theta)G(\theta)\cos\theta+F(\theta)G'(\theta)\cos\theta-2F(\theta)G(\theta)\sin\theta}{\sqrt{F(\theta)G(\theta)}}\\
&=(2F(\theta)+\cos\theta)\left(\cos2\theta-\frac{G(\theta)-F(\theta)\cos^2\theta}{\sqrt{F(\theta)G(\theta)}}\sin\theta\right).
\end{align*}
Therefore, the equation $d\tilde{R}(\theta)/d\theta=0$ reduces to
\[
2F(\theta)+\cos\theta=0
\]
or
\begin{equation}
\cos2\theta-\frac{G(\theta)-F(\theta)\cos^2\theta}{\sqrt{F(\theta)G(\theta)}}\sin\theta=0.
\label{eq:condition}
\end{equation}

First, we examine $2F(\theta)+\cos\theta=2\tilde{v}_0^2-2+3\cos\theta=0$.
The condition $\tilde{v}_0^2<1$ is necessary for a solution with $\theta<\pi/2$ (i.e., $\cos\theta>0$):
\begin{equation}
\cos\theta=\frac{2}{3}(1-\tilde{v}_0^2).
\label{eq:solution_fake}
\end{equation}
When $\tilde{v}_0^2<1$, the angle $\theta$ of the pendulum is limited to $\cos\theta\ge1-\tilde{v}_0^2$.
Therefore, the solution in Eq.~\eqref{eq:solution_fake} cannot be realized.

From the above argument, the angle $\theta$ that maximizes $\tilde{R}(\theta)$ satisfies Eq.~\eqref{eq:condition}.
The square root can be eliminated by suitably squaring Eq.~\eqref{eq:condition}:
\begin{equation}
F(\theta)G(\theta)\cos^2 2\theta=(G(\theta)-F(\theta)\cos^2\theta)^2\sin^2\theta.
\label{eq:FGcos}
\end{equation}
The right-hand side can be further manipulated as
\begin{align*}
&(G(\theta)-F(\theta)\cos^2\theta)^2\sin^2\theta \\
&=(\tilde{H}+\tilde{v}_0^2-2F(\theta)\cos^2\theta)^2\sin^2\theta\\
&=(\tilde{H}+\tilde{v}_0^2)^2\sin^2\theta-4(\tilde{H}+\tilde{v}_0^2)F(\theta)\cos^2\theta\sin^2\theta+4F(\theta)^2\cos^4\theta\sin^2\theta\\
&=(\tilde{H}+\tilde{v}_0^2)^2\sin^2\theta-F(\theta)G(\theta)\sin^22\theta.
\end{align*}
Equation~\eqref{eq:FGcos} is simplified to
\begin{equation}
F(\theta)G(\theta)=(\tilde{H}+\tilde{v}_0^2)^2\sin^2\theta.
\label{eq:FG}
\end{equation}
Substituting back $F(\theta)$ and $G(\theta)$ in Eq.~\eqref{eq:FGdef} into Eq.~\eqref{eq:FG}, we obtain
\[
(\tilde{H}+\tilde{v}_0^2)^2\sin^2\theta+(\tilde{v}_0^2-1+\cos\theta)(\tilde{H}+\tilde{v}_0^2)-(\tilde{v}_0^2-1+\cos\theta)^2\cos^2\theta=0.
\]
This equation still appears complicated, but the left-hand side can be factored relatively easily by treating it as a quadratic in $(\tilde{H}+\tilde{v}_0^2)$:
\[
(\tilde{H}+1-\cos\theta)[\cos^3\theta+(\tilde{H}+2\tilde{v}_0^2-1)\cos^2\theta-(\tilde{H}+\tilde{v}_0^2)]=0.
\]
The equation $\tilde{H}+1-\cos\theta=0$ is not suitable, since it has no solution with $\cos\theta\le1$ when $\tilde{H}>0$, and only the solution $\cos\theta=1$ ($\theta=0$) when $\tilde{H}=0$.
Thus, the remaining factor determines $\theta_\mathrm{max}$:
\begin{equation}
\cos^3\theta_\mathrm{max}+(\tilde{H}+2\tilde{v}_0^2-1)\cos^2\theta_\mathrm{max}-(\tilde{H}+\tilde{v}_0^2)=0,
\label{eq:cubic_cos}
\end{equation}
which is a cubic equation in $\cos\theta_\mathrm{max}$.

Equation~\eqref{eq:cubic_cos} is the governing equation of $\theta_\mathrm{max}$ and is analyzed in detail in the present study.
A similar equation was presented by Bittel~\cite{Bittel2005}; however, its derivation process was not provided and its theoretical treatments were not fully explored.
One contribution of the present study is the derivation of the cubic equation~\eqref{eq:cubic_cos} for $\cos\theta_\mathrm{max}$ efficiently by introducing auxiliary functions $F(\theta)$ and $G(\theta)$.

Following Bittel~\cite{Bittel2005}, a brief letter by Mungan~\cite{Mungan2014} proposed applying Cardano's formula to Bittel's cubic equation, without presenting the explicit solution.
Cardano's formula~\cite{Tignol} is certainly useful for computing the solution $\cos\theta_\mathrm{max}$ when the values of $\tilde{v}_0$ and $\tilde{H}$ are specified.
However, its expression is too lengthy to develop theoretical investigation.
This formula expresses three solutions of a cubic equation in terms of cube roots of 1 and one-third-power operation, making it difficult to identify which real and complex solutions.
Moreover, Cardano's formula does not directly existence the unique existence of solutions satisfying $0\le\cos\theta_\mathrm{max}\le1$ for any $\tilde{v}_0>0$ and $\tilde{H}\ge0$.
Therefore, applying Cardano's formula suggested by Mungan~\cite{Mungan2014} is not well suited for theoretical analysis of Eq.~\eqref{eq:cubic_cos}.
Instead, the present study focuses on analyzing this cubic equation using alternative techniques to clarify the properties of $\theta_\mathrm{max}$ and to connect them with the behavior of the maximum range $\tilde{R}(\theta_\mathrm{max})$.

\subsection{Analysis of the optimal angle $\theta_\mathrm{max}$}
Let $x=\cos\theta_\mathrm{max}$ and define $f(x)=x^3+(\tilde{H}+2\tilde{v}_0^2-1)x^2-(\tilde{H}+\tilde{v}_0^2)$.
Then, Eq.~\eqref{eq:cubic_cos} can be written as $f(x)=0$.
From $f(1)=2\tilde{v}_0^2>0$ and
\[
f\left(\frac{1}{\sqrt{2}}\right)=-\frac{2-\sqrt{2}}{4}-\frac{\tilde{H}}{2}<0,
\]
the intermediate value theorem guarantees the existence of at least one solution of $f(x)=0$ in the interval $1/\sqrt{2}<x<1$, corresponding to $0<\theta_\mathrm{max}<\pi/4$.
Although a cubic equation may have up to three (real) solutions, it can be shown that $f(x)=0$ has exactly one solution in the interval $1/\sqrt{2}<x<1$ as follows.
Consider the derivative
\[
f'(x)=(3x^2-2x)+2(\tilde{H}+2\tilde{v}_0^2)x.
\]
For $1/\sqrt{2}<x<1$, both terms $(3x^2-2x)$ and $2(\tilde{H}+2\tilde{v}_0^2)x$ are positive, implying that $f'(x)>0$ in this interval (note that $3x^2-2x$ is positive if $x>2/3$).
Therefore, $f(x)$ is strictly increasing, and the solution in this interval is unique for any $\tilde{H}\ge0$ and $\tilde{v}_0>0$.

The function $f(x)$ can also be expressed as
\[
f(x)=x^2(x-1)+(x^2-1)\tilde{H}+(2x^2-1)\tilde{v}_0^2,
\]
and each term on the right-hand side is negative in the interval $0\le x\le 1/\sqrt{2}$ for any $\tilde{H}\ge0$ and $\tilde{v}_0>0$.
Consequently, $f(x)=0$ has no solutions in $0\le x\le1/\sqrt{2}$ (i.e., $\pi/4\le\theta\le\pi/2$).

\begin{figure}[t]\centering
\includegraphics[scale=0.8]{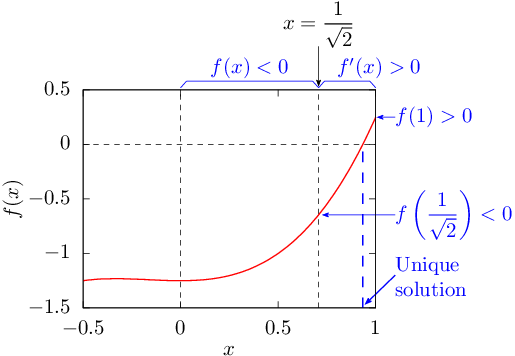}
\caption{
Existence of a unique positive solution of $f(x)=0$.
This graph corresponds to $\tilde{H}=1$ and $\tilde{v}_0=0.5$.
Opposite signs of $f(1/\sqrt{2})$ and $f(1)$ indicate the existence of at least one solution in $1/\sqrt{2}<x<1$, whose uniqueness is guaranteed given that $f'(x)>0$ in this interval.
In $0<x<1/\sqrt{2}$, no solutions exist because $f(x)<0$.
}
\label{fig2}
\end{figure}

Figure~\ref{fig2} illustrates the unique existence of the solution of $f(x)=0$ in the interval $0<x<1$.
The graph shows $f(x)$ for $\tilde{H}=1$ and $\tilde{v}_0=0.5$, although the theoretical result regarding the existence and uniqueness holds for all $\tilde{H}\ge0$ and $\tilde{v}_0>0$.

The existence and uniqueness of a positive solution of $f(x)=0$ can be shown more concisely using Descartes' rule of signs~\cite{Anderson1998}.
The nonzero coefficients of $f(x)$, in descending order, are $1(>0)$, $\tilde{H}+2\tilde{v}_0^2-1$, and $-(\tilde{H}+\tilde{v}_0^2)(<0)$.
Regardless of whether $\tilde{H}+2\tilde{v}_0^2-1$ is positive, negative, or 0, the number of sign changes between consecutive coefficients is one.
Descartes' rule of signs guarantees that $f(x)$ has exactly one positive root for any $\tilde{H}$ and $\tilde{v}_0$.

We now derive the asymptotic form of $\theta_\mathrm{max}$ for $\tilde{v}_0\ll 1$.
In this limit, the maximum angular displacement of the pendulum is small, and thus, $\theta_\mathrm{max}$ must also be small.
In particular, as $\tilde{v}_0\to0$, the solution of $f(x)=2x^3+2(\tilde{H}-1)x^2-2\tilde{H}=0$ is $x=1$ (i.e., $\theta_\mathrm{max}=0$).
Assuming $x=1-c\tilde{v}_0^2+o(\tilde{v}_0^2)$ and solving $f(x)=0$ up to order $\tilde{v}_0^2$, the coefficient $c$ is found to be
\[
c=\frac{1}{2\tilde{H}+1}.
\]
Using $x\simeq1-c\tilde{v}_0^2$ together with $x=\cos\theta_\mathrm{max}\simeq 1-\theta_\mathrm{max}^2/2$, the optimal angle is approximated as
\begin{equation}
\theta_\mathrm{max}\simeq\sqrt{\frac{2}{2\tilde{H}+1}}\tilde{v}_0=\frac{v_0}{\sqrt{g(2H+L)}}.
\label{eq:thetamax_low}
\end{equation}
Thus, $\theta_\mathrm{max}$ is proportional to $\tilde{v}_0$ (and hence to $v_0$) in the low-initial-velocity limit.

Next, we derive the asymptotic form of $\theta_\mathrm{max}$ for $\tilde{v}_0\gg 1$.
In this regime, the dominant terms in $f(x)$ are $f(x)\simeq2\tilde{v}_0^2x^2-\tilde{v}_0^2$, and the solution of $f(x)=0$ is $x=1/\sqrt{2}$, corresponding to $\theta_\mathrm{max}=\pi/4$.
This limiting angle of \ang{45} for large $v_0$ was previously observed numerically by Mungan and Lipscombe~\cite{Mungan2011} and Rave and Sayers~\cite{Rave2013}.
In this limit, the effects of the pendulum and height $\tilde{H}$ become negligible, and the situation effectively reduces to projectile motion from ground level.

To examine the asymptotic behavior of $\theta_\mathrm{max}$ to $\pi/4$, we set
\[
x=\frac{1}{\sqrt{2}}+\frac{C}{\tilde{v}_0^2}+o(\tilde{v}_0^{-2}),
\]
where $C$ is a constant to be determined.
Substituting into $f(x)=0$ and retaining terms up to order $O(\tilde{v}_0^{-2})$, we obtain
\[
C=\frac{\tilde{H}+1}{4\sqrt{2}}-\frac{1}{8}.
\]
This quantity is positive for all $\tilde{H}\ge0$.
Therefore,
\begin{equation}
\theta_\mathrm{max}\simeq\arccos\left(\frac{1}{\sqrt{2}}+\frac{C}{\tilde{v}_0^2}\right)\simeq\frac{\pi}{4}-\frac{1}{4\tilde{v}_0^2}\left(\tilde{H}+1-\frac{1}{\sqrt{2}}\right)
=\frac{\pi}{4}-\frac{g}{2v_0^2}\left(H+\left(1-\frac{1}{\sqrt{2}}\right)L\right).
\label{eq:thetamax_high}
\end{equation}

We now show that $\theta_\mathrm{max}$ increases monotonically with $\tilde{v}_0$.
Treating $\theta_\mathrm{max}$ as a function of $\tilde{v}_0$ and differentiating Eq.~\eqref{eq:cubic_cos} with respect to $\tilde{v}_0$, we obtain
\[
\frac{d\theta_\mathrm{max}}{d\tilde{v}_0}
=\frac{4\tilde{v}_0\cos2\theta_\mathrm{max}}{\sin2\theta_\mathrm{max}(3\cos\theta_\mathrm{max}-2+4\tilde{v}_0^2+2\tilde{H})}.
\]
For $\theta_\mathrm{max}<\pi/4$, we have $d\theta_\mathrm{max}/d\tilde{v}_0>0$ because
\[
3\cos\theta_\mathrm{max}-2>3\cos\frac{\pi}{4}-2=\frac{3}{\sqrt{2}}-2>0.
\]
Moreover, $d\theta_\mathrm{max}/d\tilde{v}_0=0$ when $\theta_\mathrm{max}=\pi/4$.
Therefore, $\theta_\mathrm{max}$ increases monotonically with $\tilde{v}_0$, approaching $\pi/4$.

\begin{figure}[t]\centering
\raisebox{38mm}{(a)}\hspace{-1mm}
\includegraphics[scale=0.8]{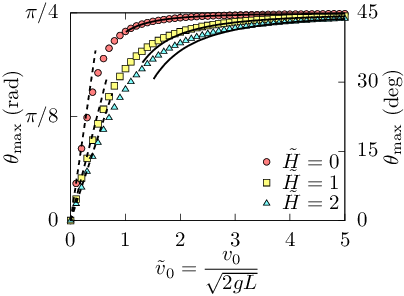}
\hspace{2mm}
\raisebox{38mm}{(b)}\hspace{-1mm}
\includegraphics[scale=0.8]{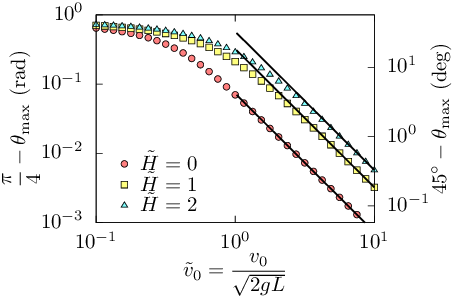}
\caption{
(a) Relation between $\theta_\mathrm{max}$ and $\tilde{v}_0=v_0/\sqrt{2gL}$ for $\tilde{H}=0$ (circles), 1 (squares), and 2 (triangles).
$\theta_\mathrm{max}$ converges monotonically to $\pi/4=\ang{45}$ as $\tilde{v}_0\to\infty$.
The dashed lines and solid curves indicate the asymptotic forms for $\tilde{v}_0\ll 1$ [Eq.~\eqref{eq:thetamax_low}] and $\tilde{v}_0\gg1$ [Eq.~\eqref{eq:thetamax_high}], respectively.
(b) The difference between $\theta_\mathrm{max}$ and its limiting value $\pi/4$ as a function of $\tilde{v}_0$ on a log-log scale for $\tilde{H}=0$ (circles), 1 (squares), and 2 (triangles).
The solid lines indicate the asymptotic power-law function~\eqref{eq:thetamax-45}.
}
\label{fig3}
\end{figure}

Figure~\ref{fig3}(a) shows the relationship between $\theta_\mathrm{max}$ and $\tilde{v}_0$ for $\tilde{H}=0$, $1$, and $2$.
The values of $\theta_\mathrm{max}$ were obtained numerically by solving the cubic equation $f(x)=0$ using Newton's method with the initial guess $x=1$.
All numerical calculations in this study were performed using Python.
The result indicate that $\theta_\mathrm{max}$ increases with $\tilde{v}_0$ for each value of $\tilde{H}$, with a slower rate of increase for larger $\tilde{H}$.
The dashed lines near $\tilde{v}_0=0$ correspond to Eq.~\eqref{eq:thetamax_low}, whereas the solid curves correspond to Eq.~\eqref{eq:thetamax_high}.
To further examine the convergence toward $\pi/4$ as $\tilde{v}_0\to\infty$, Fig.~\ref{fig3}(b) presents $\pi/4-\theta_\mathrm{max}$ as a function of $\tilde{v}_0$ on logarithmic scales.
From Eq.~\eqref{eq:thetamax_high}, the asymptotic behavior is given by
\begin{equation}
\theta_\mathrm{max}-\frac{\pi}{4}\simeq\frac{1}{4}\left(\tilde{H}+1-\frac{1}{\sqrt{2}}\right)\tilde{v}_0^{-2}.
\label{eq:thetamax-45}
\end{equation}
The numerical results are in good agreement with this power law behavior, exhibiting an exponent of $-2$ in the large-$\tilde{v}_0$ regime.

\subsection{Analysis of the maximum range}
Based on the results obtained for $\theta_\mathrm{max}$, we now analyze the maximum range $R(\theta_\mathrm{max})$.

Using Eqs.~\eqref{eq:R_FG} and \eqref{eq:FG}, $\tilde{R}(\theta_\mathrm{max})$ can be expressed as
\begin{align}
\tilde{R}(\theta_\mathrm{max})&=\sin\theta_\mathrm{max}+2F(\theta_\mathrm{max})\sin\theta_\mathrm{max}\cos\theta_\mathrm{max}+2\cos\theta_\mathrm{max}(\tilde{H}+\tilde{v}_0^2)\sin\theta_\mathrm{max}\nonumber\\
&=\sin\theta_\mathrm{max}[1-2\cos\theta_\mathrm{max}+2\cos^2\theta_\mathrm{max}+2\tilde{H}\cos\theta_\mathrm{max}+4\tilde{v}_0^2\cos\theta_\mathrm{max}],
\label{eq:Rmax}
\end{align}
which is much simpler than Eq.~\eqref{eq:R} including a square root.
Note that Eq.~\eqref{eq:FG} holds only when $\theta=\theta_\mathrm{max}$, and thus, this expression for $\tilde{R}(\theta_\mathrm{max})$ is not correct for other $\theta\ne\theta_\mathrm{max}$.

For $\tilde{v}_0\ll 1$, the following approximations are obtained from Eq.~\eqref{eq:thetamax_low}:
\[
\sin\theta_\mathrm{max}\simeq\theta_\mathrm{max}\simeq\sqrt{\frac{2}{2\tilde{H}+1}}\tilde{v}_0
\]
and
\[
1-\cos\theta_\mathrm{max}\simeq\frac{\theta_\mathrm{max}^2}{2}\simeq\frac{\tilde{v}_0^2}{2\tilde{H}+1}.
\]
Retaining only the leading-order terms of Eq.~\eqref{eq:Rmax}, which are of order $O(\tilde{v}_0)$, we obtain
\begin{equation}
\tilde{R}(\theta_\mathrm{max})\simeq\sqrt{2(2\tilde{H}+1)}\tilde{v}_0.
\label{eq:tildeRmax_low}
\end{equation}
The corresponding dimensional form (without tildes) is
\begin{equation}
R(\theta_\mathrm{max})\simeq\sqrt{\frac{2H+L}{g}}v_0.
\label{eq:Rmax_low}
\end{equation}

For $\tilde{v}_0\gg1$, the leading-order term of $\tilde{R}(\theta)$ is of $O(\tilde{v}_0^2)$, yielding
\[
\tilde{R}(\theta_\mathrm{max})\simeq4\tilde{v}_0^2\sin\theta_\mathrm{max}\cos\theta_\mathrm{max}=2\tilde{v}_0^2,
\]
where $\theta_\mathrm{max}\simeq\pi/4$ is used in the last equality.
The corresponding dimensional form is
\[
R(\theta_\mathrm{max})\simeq\frac{v_0^2}{g}.
\]
This expression coincides with the maximum range for a projectile launched from ground level.
This agreement is consistent with $\theta_\mathrm{max}\simeq\pi/4$ obtained in the previous subsection, where the effects of pendulum and the height $H$ become negligible in the leading-order approximation for $\tilde{v}_0\gg1$.

To improve the approximation, a more detailed calculation of $\tilde{R}(\theta_\mathrm{max})$ up to terms of order $O(\tilde{v}_0^0)$ yields
\begin{equation}
\tilde{R}(\theta_\mathrm{max})= 2\tilde{v}_0^2+\tilde{H}+\sqrt{2}-1+o(\tilde{v}_0^0),
\label{eq:tildeRmax_high}
\end{equation}
whose dimensional form is
\begin{equation}
R(\theta_\mathrm{max})\simeq \frac{v_0^2}{g}+H+(\sqrt{2}-1)L.
\label{eq:Rmax_high}
\end{equation}

\begin{figure}[t]\centering
\includegraphics[scale=0.8]{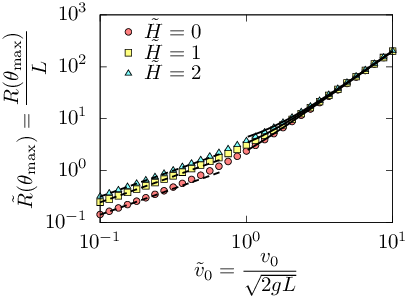}
\caption{
Maximum range $\tilde{R}(\theta_\mathrm{max})$ as a function of $\tilde{v}_0$ on a log--log scale for $\tilde{H}=0$ (circles), 1 (squares), and 2 (triangles).
The dashed and solid curves correspond to the asymptotic forms given by Eqs.~\eqref{eq:Rmax_low} and \eqref{eq:Rmax_high}, respectively.
}
\label{fig4}
\end{figure}

From Eqs.~\eqref{eq:Rmax_low} and \eqref{eq:Rmax_high}, the maximum range $R(\theta_\mathrm{max})$ is proportional to $v_0$ in the regime $v_0\ll 1$, whereas it is proportional to $v_0^2$ for $v_0\gg1$.
Figure~\ref{fig4} presents numerical results for $\tilde{R}(\theta_\mathrm{max})$ with $\tilde{H}=0$ (circles), 1 (squares), and 2 (triangles).
The dashed and solid curves correspond to the asymptotic expressions~\eqref{eq:tildeRmax_low} for $\tilde{v}_0\ll 1$ and \eqref{eq:tildeRmax_high} for $\tilde{v}_0\gg1$, respectively.

\subsection{Analysis without using cubic equation for $\theta_\mathrm{max}$}
Thus far, several results have been obtained by analyzing the cubic equation~\eqref{eq:cubic_cos} for $\cos\theta_\mathrm{max}$ intensively.
In this subsection, we investigate $\theta_\mathrm{max}$ and $\tilde{R}(\theta_\mathrm{max})$ without relying on this equation.

\begin{figure}[t]\centering
\newlength{\widefigwidth}
\setlength{\widefigwidth}{110mm}
\begin{minipage}{\widefigwidth}
\raggedright
\raisebox{27.3mm}{(a)}\includegraphics[scale=1.0]{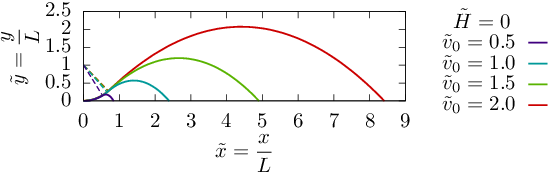}\\
\raisebox{30.4mm}{(b)}\includegraphics[scale=1.0]{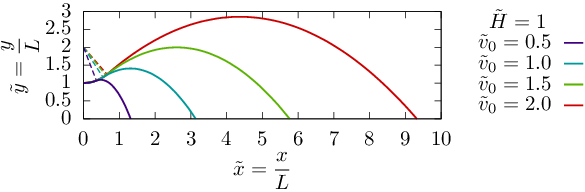}\\
\raisebox{36.1mm}{(c)}\includegraphics[scale=1.0]{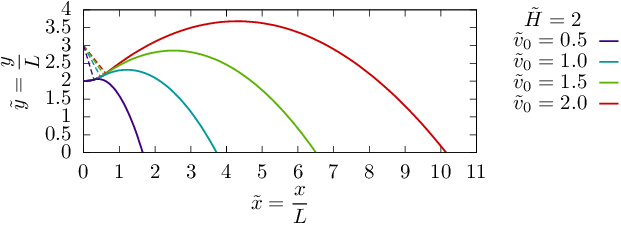}
\end{minipage}
\caption{
Optimal trajectories achieving $\tilde{R}(\theta_\mathrm{max})$ for $\tilde{H}=0$, 1, and 2, corresponding to panels (a), (b), and (c), respectively.
$\tilde{v}_0=0.5$, $1.0$, $1.5$, and $2.0$.
Dashed lines indicate release points at $\theta_\mathrm{max}$.
}
\label{fig5}
\end{figure}

Figure~\ref{fig5} shows optimal trajectories that achieve the maximum horizontal range $\tilde{R}(\theta_\mathrm{max})$ for $\tilde{H}=0$, 1, and 2, with non-dimensional coordinate $(\tilde{x}, \tilde{y})=(x, y)/L$.
Each panel contains trajectories for $\tilde{v}_0=0.5$, $1.0$, $1.5$, and $2.0$.

Trajectories for large $\tilde{v}_0$ are mostly parabolic, except for a relatively small initial portion corresponding to the pendulum motion.
The effect of the pendulum is small for large $\tilde{v}_0$, and hence, $\tilde{R}(\theta_\mathrm{max})\simeq 2\tilde{v}_0^2$ or $R(\theta_\mathrm{max})\simeq v_0^2/g$.
This is a visual evidence for $\tilde{R}(\theta_\mathrm{max})\propto\tilde{v}_0^2$ in the large-$\tilde{v}_0$ regime.

In contrast, in trajectories for small $\tilde{v}_0$, the pendulum motion makes a non-negligible contribution.
In particular, the angle $\theta$ of the pendulum is limited to $O(\tilde{v}_0)$ for small $\tilde{v}_0$.
More specifically, the conservation of mechanical energy yields
\[
\cos\theta\ge 1-\tilde{v}_0^2,
\]
and by using $\tilde{v}_0\ll 1$,
\[
\theta \le \sqrt{2}\tilde{v}_0.
\]
Therefore, the possible release angle $\theta$ can be expressed as $\theta=\sqrt{2}\alpha\tilde{v}_0$ with $0\le\alpha\le1$. %$\theta_\mathrm{max}$ is at most $O(\tilde{v}_0)$, and we reasonably put $\theta_\mathrm{max}=\sqrt{2}\alpha\tilde{v}_0$ with $0<\alpha<1$.
As stated in Subsection~\ref{sec3.1}, the speed of the object at $\theta$ is $v=\sqrt{v_0^2-2gL(1-\cos\theta)}$, whose nondimensional form becomes $\tilde{v}=\sqrt{\tilde{v}_0^2-(1-\cos\theta)}\simeq\sqrt{1-\alpha^2}\tilde{v}_0$.
The time $T_\mathrm{flight}$ from release to landing, given in Eq.~\eqref{eq:Tflight} can be approximated as $T_\mathrm{flight}\simeq\sqrt{2H/g}$.
This leading term of $T_\mathrm{flight}$ is equal to the landing time of the object launched horizontally from the initial position.
The nondimensional form of $T_\mathrm{flight}$ becomes
\[
\tilde{T}_\mathrm{flight}=\frac{T_\mathrm{flight}}{\sqrt{L/(2g)}}\simeq2\sqrt{\tilde{H}}.
\]
Here, since the characteristic length and velocity adopted in this study are $L$ and $\sqrt{2gL}$, respectively (see Eqs.~\eqref{eq:nondim_length} and \eqref{eq:nondim_velocity}), the characteristic time scale is given by $L/\sqrt{2gL}=\sqrt{L/(2g)}$.
Therefore, the nondimensional horizontal range for small $\tilde{v}_0$ is expressed as
\[
\tilde{R}(\theta)=\sin\theta+\tilde{T}_\mathrm{flight}\tilde{v}\cos\theta
\simeq \left[\sqrt{2}\alpha + 2\sqrt{(1-\alpha^2)\tilde{H}}\right]\tilde{v}_0.
\]
This expression explicitly demonstrates a trade-off situation, mentioned in Section~\ref{sec2} as ``Tarzan's dilemma''~\cite{Rave2013}.
Large $\alpha$ increases the first term $\sqrt{2}\alpha\tilde{v}_0$ yet decreases the second term $2\sqrt{(1-\alpha^2)\tilde{H}}$, and vice versa.
(The first and second terms represent the horizontal distance traveled during the pendulum and parabolic motions, respectively.)
The optimal $\alpha$ is characterized by
\[
\frac{d\tilde{R}(\theta)}{d\theta}=\left[\sqrt{2}+\frac{2\sqrt{\tilde{H}}\alpha}{\sqrt{1-\alpha^2}}\right]=0,
\]
that is,
\[
\alpha_\mathrm{max}=\sqrt{\frac{1}{2\tilde{H}+1}}.
\]
Therefore,
\[
\theta_\mathrm{max}=\sqrt{2}\alpha_\mathrm{max}\tilde{v}_0=\sqrt{\frac{2}{2\tilde{H}+1}}\tilde{v}_0,
\]
which is the same result as in Eq.~\eqref{eq:thetamax_low}.

Thus, the crossover between $\tilde{R}(\theta_\mathrm{max})\propto \tilde{v}_0$ for small $\tilde{v}_0$ and $\tilde{R}(\theta_\mathrm{max})\propto \tilde{v}_0^2$ for large $\tilde{v}_0$, shown in Fig.~\ref{fig4}, stems from different mechanisms in both limits.
The motion for large $\tilde{v}_0$ can be reduced to simple projectile motion, where the effect of the pendulum and height $H$ is negligibly small.
In contrast, for small $\tilde{v}_0$, the pendulum and parabolic motions contribute comparably to $\tilde{R}(\theta)$, and their trade-off determines $\theta_\mathrm{max}$.

\section{Discussion and concluding remarks}
This study presents a theoretical analysis of the motion of a projectile released from a pendulum, which has previously been studied only numerically or experimentally.
Despite the apparent simplicity of the combination of pendulum motion and parabolic trajectory, as demonstrated in this work, this problem possesses nontrivial complexity.
The optimal angle $\theta_\mathrm{max}$ is determined by the cubic equation~\eqref{eq:cubic_cos} for $\cos\theta_\mathrm{max}$.
We have shown that this equation has a unique solution satisfying $\theta_\mathrm{max}$, in contrast to simple projectile motion where $\theta_\mathrm{max}=\ang{45}$.

This study has performed the asymptotic analysis of the cubic equation~\eqref{eq:cubic_cos} for small and large $\tilde{v}_0$.
The asymptotic analysis of cubic equations arises in nonlinear phenomena such as the Duffing~\cite{Kovacic2011} and Ginzburg--Landau equations~\cite{Efremidis2000}.
The appearance of cubic equations in these contexts is natural, as these systems explicitly include cubic nonlinear terms.
In contrast, the emergence of a cubic equation in the Tarzan jump problem is unexpected because neither pendulum motion nor parabolic motion is typically associated with cubic equations.

In the regime $\tilde{v}_0^2\gg\tilde{H}\gg1$, Eqs.~\eqref{eq:thetamax_high}, \eqref{eq:tildeRmax_high}, and \eqref{eq:Rmax_high} reduce to
\[
\theta_\mathrm{max}\simeq\frac{\pi}{4}-\frac{\tilde{H}}{4\tilde{v}_0^2},\quad
\tilde{R}(\theta_\mathrm{max})\simeq 2\tilde{v}_0^2+\tilde{H},
\]
or, in dimensional form,
\begin{equation}
\theta_\mathrm{max}\simeq\frac{\pi}{4}-\frac{gH}{2v_0^2},\quad
R(\theta_\mathrm{max})\simeq\frac{v_0^2}{g}+H.
\label{eq:discussion}
\end{equation}
The condition $\tilde{v}_0^2\gg\tilde{H}$ implies that $2gH/v_0^2$ is very small.
Expanding Eqs.~\eqref{eq:thetamax_simple} and \eqref{eq:Rmax_simple}, which describe a simple launch from height $H$, up to order $O(2gH/v_0^2)$ yields the same expressions as Eq.~\eqref{eq:discussion}.
The condition $\tilde{v}_0^2\gg\tilde{H}\gg1$ is equivalent to $v_0^2/(2g)\gg H\gg L$.
Therefore, the effect of the pendulum, primarily characterized by $L$, becomes negligible, and the determination of the optimal release angle for Tarzan's jump becomes indistinguishable from that of a simple projectile launch to order $O(2gH/v_0^2)$.

As another limiting case, consider when the height $H$ is sufficiently large (i.e., $\tilde{H}\gg1,\tilde{v}_0^2$).
In this case, Eq.~\eqref{eq:cubic_cos} reduces to $\tilde{H}(\cos^2\theta_\mathrm{max}-1)=0$, which implies $\theta_\mathrm{max}=0$.
Therefore, when $H$ is large, ensuring sufficient horizontal velocity is advantageous for maximizing a long distance.
Returning to the Tarzan jump, if Tarzan jumps from a very high place, he can reach farther by simply leaping horizontally rather than grasping a rope.

A realistic extension of the present study is to incorporate air resistance.
However, even for the simplest form of air resistance proportional to the speed, i.e., viscous resistance, the pendulum motion becomes non-conservative, making the problem substantially more difficult.
In fact, even determining the maximum angular displacement as a function of initial velocity $v_0$ requires solving the nonlinear damped pendulum equation, and the maximum angle can no longer be obtained directly from a simple argument based on conservation of mechanical energy.
The situation where considering air resistance significantly changes the mathematical structure of the problem is similar to that of simple projectile motion, in which the optimal angle~\eqref{eq:thetamax_simple} in the absence of air resistance is replaced by an expression involving the Lambert W function when viscous air resistance is included~\cite{Hu2012}.
Therefore, a promising direction for extending the present analysis is to incorporate weak air resistance and employ effective approximations for damped pendulum motion~\cite{Nayfeh1995, Johannessen2014}.
In this context, the framework and results of the present study, including asymptotic formulas, provide useful theoretical benchmarks and guidance for future investigations.

Even in the absence of air resistance, the Tarzan jump problem admits various extensions, such as replacing the simple pendulum with other forms of constrained motion (e.g., a cycloidal pendulum) or considering landing on an inclined surface.
Theoretical analysis remains important in such generalized settings, and we expect that the present approach, including asymptotic expansions, is applicable to them.

\begin{acknowledgments}
This study was supported by a Grant-in-Aid for Scientific Research (C) (Grant Number JP23K03264) from Japan Society for the Promotion of Science.
\end{acknowledgments}

\section*{Data availability statement}
Data sharing is not applicable to this article because no new data were created or analyzed in this study.

\section*{Declaration of Interests Statement}
The author has no conflict of interest relevant to the content of this article.

\bibliography{bibs}
\bibliographystyle{apsrev4-2}

\end{document}